\documentclass[12pt]{aastex}
\def\trp{^{\scriptscriptstyle \,T}}
\shorttitle{Anisotropic Distribution of M\,31 Satellites}
\shortauthors{Koch \& Grebel}

\usepackage{emulateapj5}

\begin{document}

\title{The Anisotropic Distribution of M\,31 Satellite Galaxies: A Polar 
Great Plane of Early-Type Companions}

\author{Andreas Koch \& Eva K.\ Grebel}
\affil{Astronomical Institute of the University of Basel, 
Department of Physics and Astronomy,  Venusstrasse 7, 
CH-4102 Binningen, Switzerland \\  {\tt koch@astro.unibas.ch}}

\begin{abstract} The highly anisotropic distribution and apparent
alignment of the Galactic satellites in polar great planes begs the
question how common such distributions are.  The satellite system of
M31 is the only nearby system for which we currently have sufficiently
accurate distances to study the three-dimensional satellite
distribution.  We present the spatial distribution of the 15 presently
known M31 companions in a coordinate system that is centered on M\,31
and aligned with its disk.  Through a detailed statistical analysis we
show that the full satellite sample describes a plane that is inclined
by $-56\degr$ with respect to the poles of M31 and that has an r.m.s.
height of 100 kpc.  With  88\% the statistical significance of this
plane is low and it is unlikely to have a physical meaning.  We note
that the great stellar stream found near Andromeda is inclined to this
plane by $7\degr$.  Most of the M31 satellites are found within $< \pm
40\degr$ of M31's disk, i.e., there is little evidence for a Holmberg
effect.  If we confine our analysis to early-type dwarfs, we find a
best-fit polar plane within $5\degr$ to $7\degr$ from the pole of M31.
This polar great plane has a statistical significance of 
99.3\% and
includes all dSphs (except for And\,II), M32, NGC 147, and PegDIG.
The r.m.s. distance of these galaxies from the polar plane is 16 kpc.
The nearby spiral M33 has a distance of only $\sim 3$ kpc from this
plane, which points toward the M81 group.   We discuss the anisotropic
distribution of M31's early-type companions in the framework of three
scenarios, namely as  remnants of the break-up of a larger progenitor,
as tracer of a prolate dark matter halo, and as tracer of collapse
along large-scale filaments.  The first scenario requires that the
break-up must have occurred at very early times and that the dwarfs
continued to form stars thereafter to account for their stellar
population content and luminosity-metallicity relation.  The third
scenario seems to be plausible especially when considering the
apparent alignment of our potential satellite filament with several
nearby groups.  The current data do not permit us to rule out any of
the scenarios.  Orbit information is needed to test the physical
reality of the polar plane and of the different scenarios in more
detail.  
\end{abstract}


\section{Introduction}

The galaxies of the Local Group are not randomly distributed, but
exhibit a number of distinct patterns.  Firstly, there is a pronounced
morphology-density relation.  Gas-poor late-type dwarf galaxies are
mainly found in close proximity to one of the two dominant spirals in
the Local Group, the Milky Way and M31.  Typically these dwarfs have
distances of less than 300 kpc from the closest spiral and comprise
dwarf elliptical (dE) and dwarf spheroidal (dSph) galaxies.  Gas-rich
early-type dwarf galaxies (primarily dwarf irregular (dIrr) galaxies,
but also so-called transition-type dIrr/dSph galaxies; see Grebel,
Gallagher, \& Harbeck 2003 for details), on the other hand, show a
less concentrated distribution and are also common at larger distances
(e.g., Fig.\ 3 in Grebel 1999 and Fig.\ 1 in Grebel 2000).  Secondly,
the satellites of the Milky Way show an anisotropic distribution in
the sense that locations around the polar axis, well away from the
Galactic plane, are preferred, resembling the Holmberg effect
(Holmberg 1969).  Thirdly, the companions of the Milky Way and some of
the outer halo globular clusters lie close to one or two polar great
planes (e.g., Lynden-Bell 1976; Kunkel \& Demers 1976; Kunkel 1979;
Lynden-Bell 1982; Majewski 1994; Fusi Pecci et al.\ 1995; Kroupa,
Theis, \& Boily 2005).  There may be additional ``streams'' comprising
only one or a few satellites and outer halo globular clusters (e.g.,
Lynden-Bell \& Lynden-Bell 1995; Palma, Majewski, \& Johnston 2002).  

It is curious that essentially all of the Milky Way satellites appear
to be located  in one or two great planes.  A number of studies showed
that the probability of such planar alignments to have occurred by
chance is very low  (e.g., Kunkel 1979; Kroupa et al.\ 2005).  Several
suggestions were put forward to explain the non-isotropic, planar
distribution of the satellites.  According to one of these scenarios,
the Galactic satellites may be remnants of one or two larger,
meanwhile disrupted galaxies and orbit the Milky Way within the great
planes defined by their original parents (e.g., Kunkel 1979,
Lynden-Bell 1982; Palma et al.\ 1992).  Whether the orbits of all
these satellites do indeed lie within the planes is at present still
unclear.  For some, the proper motions seem to agree with motion
within the plane of apparent alignment, whereas this is apparently
ruled out for other objects (e.g., Schweitzer et al.\ 1995; Dauphole
et al.\ 1996; Grebel 1997; Schweitzer, Cudworth, \& Majewski 1997;
Palma et al.\ 2002; Piatek et al.\ 2002, 2003, 2005; Dinescu et al.\
2004).  However, the uncertainties of the proper motion measurements
are at present still uncomfortably large and will have to await more
accurate measurements with forthcoming astrometric space missions such
as ESA's Gaia and NASA's SIM (Space Interferometry Mission).  Another
scenario suggests that satellites follow their massive host's dark
matter distribution.  Kang et al.\ (2005) demonstrate that in this
case satellites may exhibit planar distributions as observed for the
Milky Way satellites although they find a distribution almost
perpendicular to the stellar Galactic plane to be unexpected.
Hartwick (1996; 2000) argues that the Galaxy's dark halo has ``an
extended prolate triaxial distribution highly inclined to the Galactic
plane'', thus accounting for the satellites' polar alignment.  In a
third, related scenario Knebe et al.\ (2004) suggest that satellites
retain the alignment with the massive primary that they had when they
first fell into the group or cluster along a filament.  Zentner et
al.\ (2005) and Libeskind et al.\ (2005) point out that cold dark
matter (CDM) hierarchical structure formation scenarios lead to highly
anisotropic collapse along filaments, naturally resulting in planar
configurations aligned with the major axis of the dark matter
distribution.  Both groups share the view that the Galactic stellar
disk should be approximately perpendicular to the major axis of the
dark matter distribution, an orientation supported by recent disk
galaxy formation simulations (Navarro, Abadi, \& Steinmetz 2004),
which may provide a natural explanation also for the Holmberg effect.
All these scenarios have one idea in common:  They all suggest that
the planar alignment reflects the plane of motion of the satellites.
 
Is the Milky Way exceptional in having its satellites located in one
or two great planes?  If such alignments are common, are they
preferentially polar? If one or several of the the above scenarios
hold, then similar great planes (possibly even polar great planes)
should also be found for the satellite systems of other galaxies.  The
Holmberg effect in itself is not a sufficient criterion for the
existence of polar planes since (with the exception of the Milky Way's
surroundings) the observational evidence for it comes from the {\em
projected}\ distribution of (often very few) satellites around distant
primaries.  Furthermore, there is some debate as to whether the
Holmberg effect really exists (compare, e.g., Sales \& Lambas 2004 and
Brainerd 2005).  If great planes generally exist, this would reveal
the orbital planes of the satellite galaxies, it would help to
elucidate the origin of the satellites, and could help to understand
the accretion history of massive galaxies.  

We can investigate these questions by turning to our next closest
spiral, M31.  M31 has a satellite system that covers the same range of
distances as the Galactic satellites.  Moreover, the distances of
these satellites have been well-determined using mainly observations
with the Hubble Space Telescope (HST).   In particular, for the
majority of these satellites heliocentric distances are available that
were measured using a combination of several distance indicators such
as the tip of the red giant branch and the horizontal branch,
permitting one to derive deprojected distances of these dwarf galaxies
from M31 with some confidence (see Grebel 2000 and Grebel et al. 2003,
Table 1).  This allows us to use the three-dimensional galaxy
distribution and to search for possible planes.  

M31 is surrounded by three dE galaxies and one dwarf-sized compact
elliptical (cE; namely M32).  It has at least seven dSph companions,
four of which were only discovered and confirmed during the last few
years (Armandroff, Davies, \& Jacoby 1998, Armandroff, Jacoby, \&
Davies 1999; Karachentsev \& Karachentseva 1999; Grebel \&
Guhathakurta 1999; Zucker et al.\ 2004a; Harbeck et al.\ 2005).
Additional very faint satellites may yet to be uncovered.
Furthermore, M31 contains one dIrr and dIrr/dSph galaxy within 300
kpc, and two more such dwarfs within a radius of 500 kpc.  Altogether
there are 13 satellites known within 300 kpc and 15 satellites within
500 kpc, whose spatial distribution can be investigated.  

The first search for possible great planes in the distribution of M31
satellites was conducted by Grebel, Kolatt, \& Brandner (1999), who at
that time had primarily ground-based distance determinations at their
disposal.  They found that seven (possibly eight) out of 13 satellites
appeared to lie within $\pm 15\degr$ of a great plane around M31 with
a probability for chance alignment of $\le 5\%$.  M33 seemed to lie
near an extension of this plane.  Grebel et al.\ (1999) saw little
evidence for a Holmberg effect in the distribution of M31's
companions. 

In the current paper we carry out a more sophisticated analysis using
improved statistical tools and largely homogeneous HST distances
wherever available.  Distances derived from HST photometry are
preferred owing to their superior seeing, resolution, and depth, and
because often several distance indicators were combined in determining
the distances.  This paper is organized as follows: Sect.~2 introduces
the method used to define a native coordinate system (CS) aligned
with the host galaxy M31.  In Sect.~3 the procedure of determining the
best-fit planes and performing statistical tests is presented together
with the resulting planes and in Sect.~4 we turn to the special subgroup of
M31's early-type satellites.  Finally, Sect.~5 discusses the results in
terms of dynamical aspects of M\,31's accretion history and
cosmological sub-structure populations. Sect.~6 then summarizes our findings.

\section{The definition of a native M31 coordinate system}

In order to determine the positions of the M31 satellites relative
to M31, we define an absolute coordinate system (CS), which is
anchored to the center of M31 and which has two of its vectors lying
in the disk place of M31.  Coordinates and distances were taken from
Zucker et al.\ (2004a) and Harbeck et al.\ (2005) for And\,IX, and from
Grebel et al.\ (2003) and Grebel (2000) for the remaining galaxies.
First, each pair of J2000 equatorial coordinates ($\alpha,\delta$) was
converted into Galactic longitude and latitude, ($l,b$), and from that
three-dimensional Cartesian ($x,y,z$) positions relative to the Sun
were calculated.
\begin{eqnarray} x\,& = &\,D_{\sun}\,\cos\,b\,\cos\,l \nonumber\\ y\,&
= &\,D_{\sun}\,\cos\,b\,\sin\,l \\ z\,& = &\,D_{\sun}\,\sin\,b,
\nonumber \end{eqnarray}
where $D_{\sun}$ denotes the observed distances from the Sun (see
Table~1). This right-handed CS (eqs.\ 1) is oriented such that $x$
points towards the Galactic center and $z$ indicates the height above
the Galactic plane. 

After applying a linear translation to move the origin of this CS to
the center of M31, the CS is aligned with this galaxy by rotation
around three angles.  The first of these affine transformations
incorporates the position angle (PA) of M31.  Accounting for the
inclination of the celestial against the Galactic pole and M\,31's PA
of (37.7\,$\pm$\,0.2)\,$\degr$ (de Vaucouleurs 1958), we rotate the CS
clockwise using a transpose rotation matrix around the
$y$-axis\footnote{Since we rotate the CS rather than the coordinates
themselves, the transpose matrix has to be used.}, $R_y\trp(p)$, by
the angle $p=115.17\degr$.  In the next step, the resulting CS is
rotated  around the new $x$-axis by inclination via the matrix
$R_x\trp(i)$, where we use the canonical value for the inclination of
M31 of $i=-12.5\degr$ (de Vaucouleurs 1959). In this notation,
90$\degr$ signifies a face-on view. The minus sign arises since the
matrices are defined for clockwise rotation.  Finally, we rotate the
resulting CS, which is now coplanar with the M\,31 galactic plane,
around its respective $z$-axis by 180$\degr$ by means of
$R_z\trp(\pi)$. Thus consistent with common representations, our
CS is oriented such that $X_{M31}$ increases toward the southwest,
$Y_{M31}$ increases toward the northwest, and $Z_{M31}$ points toward
M\,31's galactic pole. 

The transformed Cartesian coordinates are thus determined from
($X_{M31},Y_{M31},Z_{M31})\trp\,=\,R_z\trp(\pi)\cdot R_x\trp(i)\cdot
R_y\trp(p)\cdot\,(x,y,z)\trp$.  The expressions for the individual
components read: 
\begin{eqnarray} X_{M31}\,&=&\,-x\cos\,p\,+\,z\sin\,p \nonumber\\
Y_{M31}\,&=&\,-y\cos\,i\,-\,x\sin\,i\,\sin\,p\,+\,z\sin\,i\,\cos\,p \\
Z_{M31}\,&=&\,\,\,\,\,y\sin\,i\,-\,x\cos\,i\,\sin\,p\,+\,z\cos\,i\,\cos\,p.\nonumber
\end{eqnarray}

A schematic diagram of the satellites' location relative to M\,31 in
this native M31 CS is shown in Fig.~1.  The uncertainties in each of
the three coordinates were derived by applying the above
transformations accounting for the uncertainties in the distances as
the only error source.  The right panel of Fig.~1 seems to suggest the
absence of an obvious Holmberg effect in the satellite distribution.
Furthermore, by eye one may be tempted to position a possible great
circle along the approximate longitudes of $+30\degr$ and $-150\degr$,
but this does not look like a very well-defined great circle.  Since
it is difficult to determine a preferential alignment of the satellite
distribution by eye, we now pursue the question of great planes
comprising all or subsets of M31's companions via a statistical
approach. 

\section{Great planes including all satellites} 

The most convenient parameterization of a plane is the Hesse form,
which describes each point within the plane in terms of the normal
vector $\mathbf{n}$ and two vectors, $\mathbf{x}$, $\mathbf{p}$, each
pointing from the origin to a point located on the plane. Then
$\mathbf{n}\cdot (\mathbf{x}\,-\,\mathbf{p})=0$ unambiguously defines
the plane.  One can determine the closest distance $D_p$ between the
origin of the CS and the plane via $D_p=\mathbf{n}\cdot\mathbf{p}$
(see also Kroupa, Theis, \& Boily 2005). Since we seek to identify
great circles or great planes,  the plane needs to intersect the
origin (i.e., the center of M31), which allows us to set $D_p$ to
zero.  Then the Hesse form can be simplified as follows:
\begin{equation} 
n_1\,X_{M31}\,+\,n_2\,Y_{M31}\,+\,n_3\,Z_{M31}\,=\,0.
\end{equation}

Here $n_i$ denotes the respective components of the plane's normal
vector\footnote{It is often convenient to give $\mathbf{n}$ in its
spherical parameterization, i.e.,
$l\,=\,\arctan(n_2\,/\,n_1)$ and
$b\,=\,\arctan(n_3/\sqrt{n_1^2+n_2^2})$.  } and ($X_{M31},Y_{M31},Z_{M31}$) is
the position vector of each satellite, as determined above.  {From}
this the distance of any point $(x_i,y_i,z_i)$ to the plane is given
by
\mbox{$d_p=(n_1\,x_i\,+\,n_2\,y_i\,+\,n_3\,z_i)\,/\,\sqrt{n_1^2\,+\,n_2^2\,+\,n_3^2}$}.
We fit the implicitly defined surface (eq.\ 3) to our data by means of
an error-weighted orthogonal distance regression (ODR) using {\sc odrpack}
(Boggs \& Rogers 1990, Boggs et al.\ 1992). Instead of minimizing the
projected distance to the plane in a given coordinate, as in a
traditional least-squares fit, ODR takes into account the
perpendicular distance to the curves to fit.  The individual
data points were weighted in the fit by the deprojected uncertainties in the 
three-dimensional positions, which were calculated from the measurement
uncertainties in the galaxies' distances. 

\subsection{The best-fit satellite plane}

The formally best-fit
plane that we obtained by performing one single ODR fit comprising the
entire sample of 15 satellites lies at a
normal vector of $l=171.2\degr$ and $b=-45.6\degr$.  However,
anticipating the statistical method in Sect.~3.3, the significance of
this plane is 84\%, corresponding to 
1.4 Gaussian $\sigma$ and we cannot reject the possibility that such 
a plane is a purely random alignment. 
If we describe the r.m.s. height of an underlying disk distribution for
$N$ satellites as $\Delta=\sqrt{(1/N)\sum_{i=1}^{N}d_p^2(i)}$, this
value is found to be 99.4\,kpc.
It is obvious that not all satellites fall onto this plane.  Outliers
can hamper the determination of a best-fit solution for the simple
reason that they are not physically associated with the underlying
population that presumably forms such a disk.
 
\subsection{Bootstrap tests of best fit planes}

When fitting a plane to a set of data points, the influence of outliers
can be overestimated and can yield significantly different results.
However, since one cannot flag any data point as an outlier {\em a
priori}, we have to use a statistical method to reliably determine a
robust solution for estimating best-fit planes.  We approach this
problem by a bootstrap test (Efron \& Tibshirani 1993). That is, we
draw any possible combination of a subsample from the satellites,
where we covered all possible sample sizes from three to all 15
companions, thus allowing us to run $\sum_{i=3}^{15}{15 \choose i}$
different tests.  For each of the 32647 possible subsamples we
performed the plane fit as described above.  The resulting
distribution of the normal vectors of the best fit planes is shown in
the top left panel of Fig.~2, where the total of all 15 companions forms
the parent sample.  Since the direction of the pole is ambiguous due
to the lack of actual orbit information, ODR cannot distinguish
between normal vectors that are simply inverted in $b$ and shifted by
180$\degr$ in $l$. These points are then assigned to the complementary
plane exhibiting the mirrored normal vector.  The distinct peak in
Fig.~2 (left panel) occurs in the direction of $l=150.8\degr$ and
$b=-56.4\degr$, which defines a best-fit plane based on a more robust
method than obtained by a single fit of all data points.  The resultant
$\Delta$ is 100.0 kpc.  It is noteworthy that this is not a polar
alignment as would be expected if the Holmberg effect occurred also in
M\,31.  

Fig.~3 (left panel) shows the location of all the M31 companions and
the great plane that was derived from this ODR fit comprising the
entire sample of 15 satellites. The diagram is shown from a viewpoint
rotated such that the great plane is seen edge-on.  This great plane
comprises all dEs, M32, and also all dIrrs except for the
transition-type dIrr/dSph galaxy PegDIG (located at a distance of
410\,kpc to M31).  

Although not used in the fits discussed above, we superposed the
location of the Andromeda Stream (McConnachie et al.\ 2003) onto the
diagram (Fig.~3).  This stream has been shown to extend to at least
4.5$\degr$ southeastward of M31.  The ten fields from McConnachie et
al.  (2003) (errorbars were omitted for clarity) are naturally
aligned with respect to each other, but are still located in a
separate plane that is inclined against the best-fit plane of our
analysis by approximately 7$\degr$.  We did not attempt to include
other features such as And NE (Zucker et al.\ 2004b), since their
three-dimensional position is less well known.  

\subsection{Statistical significance of the planes}

In order to assess the statistical significance of the previously
determined best-fit plane, we ran a number of additional tests.

First, we generated a random sample of 15 satellites, distributed  
out to the maximum distance of the observed companions.
The radial distribution of the random satellites was taken to follow a power law
with an exponent of $-2$, which is a fairly good approximation of the
actual radial distribution of the M31 companions and is also similar to 
the prediction of cosmological sub-halos (Klypin et al. 1999, Moore et al. 1999, 
Zentner \& Bullock 2003).  By means of a KS test it can be 
shown that the cumulative sample of companions is consistent with such an isothermal 
density distribution at 99.1\%, where the most likely power-law indices fall in the range 
between  $-1.6$ and $-2.3$ (see also Kroupa et al.\ 2005). 
The innermost satellite was, however, ignored in this procedure, since the central regions of the 
M31 system are known to be incompletely sampled by observations. Fig.~4 shows the radial distribution 
of the observed satellites relative to M31. 
Then the entire procedure of bootstrap-fitting planes to this random distribution 
was carried out analogously to that of the real observed set as described in Sect.~3.2. 
We determined the best-fit plane from the corresponding density maps  
(see the bottom panel of Fig.~2 for a sample) 
 of the 32647 combinations  
and calculated the respective r.m.s.\ distance of the 15 random points to this plane. 
This procedure was repeated a large number of times (of the order of 10$^3$) 
to allow us to assess the probability that the r.m.s.\ distance $\Delta$ originates 
from a random distribution and to also identify any other potential biases in our 
method.
 
For the best-fit plane to the entire satellite sample of 15 companions we find a 
significance of 
87.4\%, hence our result is robust at the 1.5$\,\sigma$-level. 
Therefore we cannot reject the hypothesis that such a plane may result 
from a random distribution and thus may not have any physical meaning. 
Including McConnachie et al.'s ten fields from the Andromeda stream
into the fit routines did not alter the location of the resulting
plane much.  For this enlarged sample we found a normal vector of
($l=148.5\degr,\,b=-53.3\degr$) with a residual r.m.s of 78\,kpc. 
However, an interpretation of this latter result should be taken 
with caution, since the sample is biased 
toward the stream due to the incorporation of ten fields for one contiguous feature
(i.e., the stream) versus 15 individual satellites. Hence, stating any significance will 
not be meaningful as we would produce an artificially increased significance from the large number of fields.

A second test for the robustness of the fitting method employed here
comprised the rotation 
of the real galaxy sample by pairs of random angles.  The resulting
data were subsequently subjected to the same 
fitting procedure as above, again repeated for a large number of samples. 
As a result, we could recover the best-fit plane rotated by the input random angles, where 
the scatter around the original best-fit angles amounts to approximately 5--10$\degr$.  
This lends further support to the results obtained with the method used here
and additionally 
provides an estimate of typical uncertainties that result from the fits.

\section{A polar plane of early-type M31 companions}

In the previous section we analyzed the entire sample of M31
companions comprising dEs, cEs, dSphs, and dIrrs as well as transition
types such as the dIrr/dSph LGS3.  Since the dSphs form the most
numerous dwarf subsample in a galaxy group, and since the majority of
satellite candidates of the massive Local Group galaxies are dSphs
(e.g., Grebel 1999), we performed the bootstrap fit procedure
including only the seven dSph satellites.  It is noteworthy that,
while a fit to the full sample of all M31 satellites does not yield a
highly significant, unambiguous solution, the majority of dSphs lies
within a plane defined by $l=107.1\degr$ and $b=6.9\degr$ (see middle
panel of Fig.~2) with an r.m.s.\ of $\Delta=46\,$kpc.  This plane is
indicated in the middle panel of Fig.~3 after rotation of the
viewpoint by the respective longitude.  Only one dSph deviates
considerably from this plane: And\,II, located at a distance of 158
kpc to M31 and 112\,kpc to the plane, where the latter value is larger
than two standard deviations.  Excluding this obvious outlier yields a
high significance (determined as above by a large number of random samples 
of seven satellites) 
of the resulting dSph plane of 99.7\%,  corresponding to 3\,$\sigma$. 

As Fig.~3 (middle panel) implies, also M31's close companions, the cE
M32 and the dE NGC\,147 (as well as the transition-type dIrr/dSph
PegDIG), lie reasonably close to the best-fit dSph plane.  We may thus
ask whether an improved fit would result when {\em all}
morphologically similar galaxies are included, i.e., all galaxies of
the dSph/dE/cE class.  We reran the bootstrap test on this enlarged
subgroup.  This procedure yields a slightly different plane at
$l=102.5\degr$ $b=5.2\degr$ (see right panels of Figs.~2 and 3) with
an r.m.s.\ of the residuals of $\Delta=51$\,kpc (without And\,II).  As
a result, the significance amounts to 98.7\% ($2.5\,\sigma$), again
excluding the outlier And\,II.  Hence there is very little difference
as compared to the previous fit that included only dSphs.
However, if we consider only those galaxies whose positions seem to be in
good agreement with the polar great plane of early-type companions, namely 
 M32, NGC 147, PegDIG and the dSphs, but excluding And\,II,
NGC\,185, and NGC\,205, these nine companions lie
within a thin disk with an r.m.s. distance of 16\,kpc to this early-type plane. 

Interestingly, also the smaller Local Group spiral M33 is directly
encompassed by this great circle (its orthogonal distance to this
plane being 2.8\,kpc).  However, it is not related to the great plane 
resulting from the fit to {\em all} M31 satellites -- here M33 has a
distance of 135\,kpc from the plane.
Moreover, while the plane comprising all of the M31 satellites is
highly non-polar (at $-56\degr$), the great plane that includes dSphs,
dEs, and the cE exhibits a nearly polar alignment with an inclination
of $5-7\degr$ from M31's pole.

The most luminous, most massive globular cluster in the Milky Way,
$\omega$ Cen, shares a number of properties with dwarf galaxies and is often
considered to be the stripped remnant of an accreted dwarf (e.g., McWilliam
\& Smecker-Hane 2005; Hilker et al.\ 2004; Ideta \& Makino 2004; and 
references therein).  The most massive, most luminous globular clusters
known in the Local Group are located in M31.  One may speculate that these
objects might be nuclei or bulges of stripped dwarfs.  G1, for instance,
also seems to exhibit a metallicity spread (Meylan et al.\
2001).  If so, the progenitors of these luminous clusters may
also have been early-type dwarfs.  We have compared the location of the
two most luminous M31 globulars, Mayall II or G1 and B\,327 (van den
Bergh 1968), to the location of our polar plane of early-type galaxies. 
Although we did not include these objects in any of the fits, we indicate  
in Figs.\ 1, 3, and 5 also the positions of these massive globular clusters 
relative to the M31 system. These clusters have an 
adopted distance coincident with that of M31 itself (Rich et al.\ 1996, 
Barmby et al.\ 2002).  While G1, which is often regarded as the most 
luminous globular cluster of the Local Group, 
lies at a distance of merely 8\,kpc of this polar great plane and 
coincides with it to within its errorbars, 
B\,327 is fully encompassed by the early-type great 
plane\footnote{Van den Bergh (1968) argues that B\,327 is probably the most
luminous globular cluster when its reddening is properly taken into account.}. 

Uncertainties in the analyses presented here result not only from
uncertainties in the distances to the satellites of M31, but also from
the uncertainty of the distance to M31 itself.  Hence we carried out
our analysis for three widely used distances to M31 from the
literature.  The results discussed above rely on the Cepheid distance
of 773 kpc (Freedman \& Madore 1990).  In addition, we also adopted
the mean M31 distance of 783 kpc based on several distance indicators
discussed by Rich et al.\ (2005), and the mean distance of 760 kpc
resulting from various distance indicators given by van den Bergh
(1999).  The formal mean uncertainties of these distance measurements
are of the order of 10--20 kpc, implying that the different distances
agree within their uncertainties.  The results for all three M31
distances are listed in Table~2.  As the values in Table~2
demonstrate, the above variation of the distance to M31 does not
significantly alter our results.  The statistical presence of a polar
great plane prevails.   

\section{Discussion}

\subsection{The break-up or tidal remnant scenario}

As mentioned in Section 1, one of the scenarios put forward to explain
the Galactic polar planes suggests that the dwarf galaxies within such
a plane are tidal remnants of a more massive galaxy (e.g., Kunkel
1979; Lynden-Bell 1982).  We will refer to this idea as ``Scenario
I''.  This scenario leads to the question of whether the properties of
the dwarfs, in particular with respect to their stellar populations,
are consistent with an origin from a single parent galaxy.  We will
consider this question first for the Milky Way companions and then for
the M31 companions. 

\subsubsection{A few musings on the Galactic polar great planes}

For the Milky Way companions it was shown that each of these dwarfs
has its own unique evolutionary history that differs from other dwarfs
even when of the same morphological type (e.g., Grebel 1997).  This
need not contradict an origin from a common, since accreted parent,
but would indicate that the separation from this parent should have
occurred very early on, followed by continued evolution of the
individual tidal fragments.  The low metallicities of the old
populations of the dSphs (see Table 1 in Grebel et al.\ 2003) indicate
that either the parent galaxy was little evolved when its disruption
occurred (supporting the view that this must have happened at ancient
times) and/or that the dSphs are tidal fragments of the outer,
metal-poor regions of the parent.  

All nearby Galactic dwarfs seem to share a common epoch of ancient
star formation that is coeval within the present-day measurement
accuracy (to within $\sim 1$ Gyr) and that is indistinguishable from
the oldest age-datable stellar populations in the Milky Way (Grebel \&
Gallagher 2004).  The SMC appears to have an old population that is
several Gyr younger than the ancient star formation episodes in the
other dwarfs and in the Milky Way, but more detailed data are still
needed for this galaxy.  For the remaining dwarfs, a common epoch of
early star formation does not necessitate a common origin, but lends
more support to such an idea than would widely differing times for the
first significant star formation.

While the {\em mean} metallicities of the old populations in the
various dwarfs tend to differ by a few tenths of a dex, all of these
galaxies also show a considerable abundance spread among their old
stars.  Neither property precludes a common origin from fairly
metal-poor regions of a putative common progenitor.  

The Galactic dwarfs follow a metallicity-luminosity relation (e.g.,
Grebel et al.\ 2003), indicating that intrinsic processes such as
their own gravitational potential and hence the ability to retain
metals played an important role in their evolution.  If these dwarf
galaxies are leftovers stripped from a larger satellite, they must
once again have been stripped at an early time and must then have
continued to form stars after this event in order to produce the
observed metallicity-luminosity relation.  Clearly, the Galactic dSphs
are quite different from more recently formed tidal dwarfs whose
departure from the metallicity-luminosity relation readily betrays
their nature (e.g., Duc \& Mirabel 1998).  

It would seem that sustaining the extended star formation histories of
the Galactic dSphs (see, e.g., Grebel \& Gallagher 2004) would be
difficult in low-mass tidal remnants without dark matter unless these
galaxies had substantially larger baryonic masses when they condensed
than the $\sim 10^5$ -- $10^6$ M$_{\odot}$ derived today from their
stellar content (see also discussion in Grebel et al.\ 2003, and for
instance the models presented by Wang et al.\ 2005 and Mashchenko et
al.\ 2005).  Other arguments against dSphs being mere tidal remnants
without dark matter include the lack of substantial line-of-sight
depth (Klessen, Grebel, \& Harbeck 2003) and extended, fairly flat
radial velocity profiles (Wilkinson et al.\ 2004).  Nonetheless, the
question of dark matter in the low-mass dSphs remains yet to be
resolved.

Whereas some merger events that are believed to have occurred several
Gyr ago have left detectable tidal streams in the Milky Way -- for
instance, Sagittarius and possibly Monoceros, Canis Major (Ibata,
Gilmore, \& Irwin 1994; Newberg et al.\ 2002, 2003; Yanny et al.\
2003; Majewski et al.\ 2003; Martin et al.\ 2004), and
Triangulum-Andromeda (Rocha-Pinto et al.\ 2004) -- no ancient event
has so far been identified that could be connected to the origin of
the polar great planes.  If these planes do indeed have a physical
meaning, this  unsatisfactory situation 
may change with the parallax and phase space information
for the vast number of Milky Way stars that will be provided by the
Gaia mission (Perryman et al.\ 2001).  It has been suggested that the
Large Magellanic Cloud may be the main part of a broken-up parent
galaxy responsible for the Magellanic Stream of dwarf galaxies and
simply has not yet merged with the Milky Way (e.g., Kunkel 1979).
Related hypotheses propose that the precursor of today's Sgr dSph may
have been deflected into its current orbit by a collision with the LMC
(Zhao 1998), and that Fornax might have been stripped of its gas by an
encounter with the Magellanic Stream, leading to the H\,{\sc i} cloud
excess along its inferred orbit (Dinescu et al.\ 2004).  

The strongest evidence in favor or against the reality of orbital
planes of dwarf galaxies will come from proper motion measurements.
At present, recent measurements reveal a complex picture.  Ursa Minor
can be ruled out as a member of the Magellanic Stream (Piatek et al.\
2005).  Fornax, excluded as a stream member by the data of Piatek et
al.\ (2002), is proposed as a likely member of the Fornax--Leo I--Leo
II--Sextans--Sculptor stream by Dinescu et al.\ (2004).  For Carina
and for Sculptor the situation appears to be ambiguous at present
(Piatek et al.\ 2003; Schweitzer et al.\ 1995).  As more and more
epochs are being added, the measurements should yield a clearer
picture in the coming years. 

\subsubsection{A few musings on the M31 polar great plane}

The low-mass dSph satellites of M31 are all metal-poor and show hints
of metallicity spreads (e.g., C\^ot\'e, Oke, \& Cohen 1999;
Guhathakurta, Reitzel, \& Grebel 2000; Harbeck et al.\ 2001).
However, unlike the Galactic dSphs the M31 dSphs are dominated by old
populations, lacking prominent intermediate-age or even young
populations (e.g., Harbeck et al.\ 2001; Harbeck, Gallagher, \& Grebel
2004).  In this sense, they show a much higher degree of homogeneity
than the Galactic dSphs.  The two elliptical dwarfs that appear to be
associated with the polar great plane show considerable enrichment,
but NGC 147's globular clusters are old and metal-poor (Da Costa \&
Mould 1988; Han et al.\ 1997), and indications of a small old and
metal-poor population were recently found in M32 (Alonso-Garc\'{\i}a,
Mateo, \& Worthey 2004).  As for the Milky Way companions, a putative
break-up that would have produced the early-type companions of M31
would need to have occurred at very early times. 

All of the gas-deficient M31 companions follow the
metallicity-luminosity relation of early-type dwarfs (Grebel et al.\
2003; Harbeck et al.\ 2005).  In the remnant scenario, this would
imply that they should have undergone further chemical evolution to
reach a state consistent with their luminosity; hence one may suggest
that the remnants should still have contained sufficient gas to
continue to form stars for a while after the break-up.  If so, then
again the  break-up must have occurred at very early times considering
the observed absence of prominent younger populations.

The cE M32 is a very interesting object in itself:  It contains a
black hole with a mass of a few times $10^6$ M$_{\odot}$ (e.g., Tonry
1984; Joseph et al.\ 2001), it interacts with M31 (King 1962; Choi,
Guhathakurta, \& Johnston 2002), and may be the remnant of a larger
elliptical galaxy (Faber 1973; Nieto \& Prugniel 1987) or the bulge of
a stripped early-type spiral galaxy (Bekki et al.\ 2001; Graham 2002).
The latter is supported by the detection of what appears to be the
remains of a disk in M32 (Graham 2002).  This raises the intriguing
possibility that M32 may be the remnant of the parent of the dwarf
galaxies located in the M31 polar great plane identified in our paper.
On the other hand, M32 may be associated with the giant stellar stream
around M31 (Ibata et al.\ 2001), since it appears to be located within
the stream; however, its velocity is quite different from that of the
stream (Ibata et al.\ 2004).  The stream itself seems to be on a
highly radial orbit passing very close to the center of M31.
Kinematic studies suggest that its progenitor may have survived until
1.8 Gyr ago (Ibata et al.\ 2004).  Considering this and that the
stream stars are metal-rich (Ibata et al.\ 2001), an immediate
association of the stream with the dSphs seems to be ruled out.  

The M31 halo differs substantially in its properties from the Galactic
halo.  The stellar halo appears to extend beyond 150 kpc (Guhathakurta
et al.\ 2005), implying that many of the dwarf satellites considered
in our present study are in fact moving through the stellar halo of
M31.  Apart from a significant old population the halo also contains
metal-rich intermediate-age populations with ages in the range of 6--8
Gyr that appear to account for $\sim 30$ \% of the stellar mass (Brown
et al.\ 2003).  With $\sim -0.5$ dex, the mean metallicity is
comparatively high (Brown et al.\ 2003; Durrell et al.\ 2004) and
exceeds that of the dSph satellites by at last one dex in [Fe/H].
Hence a once larger population of M31-dSph-like galaxies may have
contributed to the ancient halo of M31, but was not the dominant
contributor to its complex halo population structure as a whole. 

The proximity of M31's massive globular clusters G1 and B\,327 to the 
plane of early-type satellites raises
the question whether these objects should also be considered as the
remnants of nucleated dwarf galaxies (e.g., Meylan et al.\ 2001). 
If they originate from the same break-up event as the remainder of the 
dSphs, they must have undergone a different evolution.  Primarily, they 
would then seem to have been dominated by tidal stripping and harassment from their massive 
host galaxy to leave a nucleus or bulge in its present, globular-cluster-like
form.

\subsection{The prolate dark halo scenario}

As outlined in Section 1, this scenario assumes that satellites follow
the dark matter distribution of the Milky Way.  Polar great planes
would result if the dark halo is prolate, as some authors are favoring
for disk galaxies (e.g., Hartwick 2000; Navarro et al.\ 2004) 
and as has been suggested for our own Milky Way from the kinematics of  
the Sgr dwarf tidal streams (Helmi 2004).

Our
finding that most of the {\em low-mass} satellites within 300 kpc of
M31 lie within a polar great plane is consistent with this scenario
and supports that triaxial prolate dark halos may be a common
occurrence in disk galaxies.  

We note, however, that the evidence for a Holmberg effect among the
M31 satellites is weak.  The majority of the M31 satellites is found
within $|\,b_{M31}\,| < 40\deg$ of its disk (Fig.~1, right).    

It would be highly desirable to carry out similar studies also for the
satellite populations of nearby groups.  While we now have distances
for many of the satellites in these systems based on the tip of the
red giant branch from HST photometry (see Karachentsev et al.\ 2000,
2002a, 2002b, 2003a, 2003b, 2003c for details), the uncertainties of
these distances including, in particular, those to the massive
galaxies in these groups make it difficult to reliably derive the
three-dimensional galaxy distribution with sufficient accuracy for a
comparable analysis.  

\subsection{The filament scenario}

This scenario will also result in planar alignments, which would only
be polar if that is the orientation of the major axis of the dark
matter distribution.  Both this and the preceding scenario have the
advantage that they do not require a common origin of the dwarfs and
permit the presence of dark matter in the satellites.   

An interesting consequence of this scenario is that one may expect to
find additional dwarf galaxies when following the great planes out to
larger distances since the planes should trace the location of
extended cosmological filaments.  

Fig.~5 shows face-on views on M31's disk: in the left panel of
Fig.~5, we show the present-day location of several nearby galaxy
groups, represented by their brightest object with distances adopted
from Karachentsev (2005) and projected onto the plane of M31's disk.
It is interesting to note that while the M83 and Cen A groups are
located far from the polar plane spanned by Andromeda's early-type
companions, the M81 group seems to almost coincide with this great
plane (or filament?). Also the extended Sculptor group and presumably
the Canes Venatici I Cloud appear to be approximately oriented toward
the direction of M31's polar satellite plane, albeit at larger angles.
The right panel of Fig.~5 shows the immediate surroundings of M31.
The two arrows indicate the directions toward the Milky Way and M33.
Few satellites seem to lie at the far side of M31 as seen from the
Milky Way.  There is no obvious filamentary structure of M31
satellites extending toward the Milky Way, but the polar plane of
early-type companions clearly points toward M33 as we pointed out
already earlier. 

\section{Summary}

We have presented a Cartesian coordinate system that is centered on
M31 and aligned with its disk.  We calculated the positions of the
galaxies within 500 kpc in this CS.   Most (possibly all) dwarf
galaxies within this radius are likely satellites of M31.  We then
investigated the existence of possible great planes encompassing
subsets or all of the companions.  The great plane that results when
trying to account for all 15 M31 companions has low statistical
significance (84\%) and includes many outliers.  While this plane
probably has no physical meaning, interestingly the recently
discovered Andromeda Stream lies close to it and is inclined with
respect to it by $\sim 7\degr$.  

If we restrict our sample selection to only gas-deficient galaxies,
i.e., to the dSph, dE, and the cE companions of M31, a {\em polar}\
great plane with a statistical significance of 98\% results.  This
supports the earlier claim of the existence of such a plane by Grebel
et al.\ (1999), now based on better and more comprehensive data.  M32,
NGC 147, PegDIG, and even M33's position are consistent with this
great plane.  When excluding three deviating early-type dwarfs
(And\,II, NGC\,185, and NGC\,205) as outliers from the calculation of
the statistical significance, the remaining early-type galaxies lie
within a mere 16\,kpc of this plane, and the resulting
statistical significance is 99.7\% (3$\sigma$).  The plane resembles the polar
great planes of satellites found around the Milky Way and includes
also the more distant dIrr/dSph transition-type galaxy PegDIG and even
M33.  In total, this polar plane comprises nine out of 15 M31
companions including eight out of 11 of its early-type dwarfs.  
We note that also the two most luminous globular clusters in the Local
Group, both of which are located in M31,
are coincident with the plane of early-type companions.

While the plane comprising all of the M31 satellites is clearly
non-polar (at $-56\degr$), the great plane of gas-deficient satellites
shows a nearly polar alignment with an inclination of $6-8\degr$ from
M31's pole.  We note that, in contrast to the Milky Way, the M31
companions show little evidence for a Holmberg effect.  The majority
of these companions is found within $\pm 40\degr$ of M31's equator.
Our findings are relatively insensitive to the adopted distance to M31
itself.

Several scenarios have been suggested to explain the existence of
polar planes.  A popular scenario suggests that planes originate from
the break-up of larger galaxies, keeping smaller fragments in the
orbit defined by the progenitor.  The fragments may be pure tidal
remnants devoid of dark matter.  We argue that based on the stellar
populations and metallicities of both the Milky Way and the M31
satellites, such a break-up would have to have occurred very early on.
A suitable parent progenitor yet needs to be identified.  Since the
satellites follow the luminosity-metallicity relation, they must have
continued to form stars after the break-up.  There is little evidence
so far that the satellites are devoid of dark matter as one would
expect from unbound tidal debris.  Obviously the best test of this
scenario is via proper motions and orbits.  The available proper
motions for Galactic dwarfs have disproved the association of certain
dwarfs with polar orbital planes, but may support this for others.

The prolate dark halo scenario proposes that satellites follow the
dark matter distribution of the massive galaxy they are orbiting,
requiring prolate dark halos to create polar great planes.  The
existence of polar great planes of satellites not only around the
Milky Way, but also around M31 would seem to support this scenario,
but as noted earlier there is little evidence for a pronounced
Holmberg effect in the satellite system of M31.  Ultimately, again
proper motions and orbits will provide the best test of whether the
planar alignments are fortuitous or physical.

The filament scenario suggests that satellites are oriented along
cosmological filaments of dark and baryonic matter that is gradually
accreted by massive primaries as these continue to grow in
hierarchical structure formation.  In this case planar alignments not
only in the immediate vicinity of massive galaxies are expected, but
such filaments should extend over much larger scales.  Indeed our
polar great plane of M31 satellites points toward the M81 group.  On
larger scales and for more distant galaxies, this scenario can be
statistically tested via weak lensing measurements and large galaxy
surveys (e.g., Zentner et al.\ 2005).

A clear distinction between the different scenarios is not yet
possible at present.  We can impose constraints based on the known
stellar populations and chemical properties of the satellites as
discussed before. However, we also need to keep an open mind regarding
other possibilities such as that interactions and encounters between
companion galaxies may have deflected some of them and altered their
orbits, or that we are reading too much into potentially fortuitous
planes that may be unconnected with any physical motion of the
satellites.  All in all, our study underlines the urgent need for
orbital information, part of which may be provided by future
astrometric missions.  Clearly, the distribution and motion of
satellites provides important tests of galaxy formation and evolution,
of the importance of accretion events, of the origin and nature of
dwarf galaxies, and of CDM scenarios. 

\acknowledgments

We are grateful to H.\ Jerjen, M.\ Metz, and O.\ Gerhard for helpful 
discussions.  We also thank S.\ van den Bergh and an anonymous
referee for helpful comments.  
We acknowledge support from the Swiss National Science Foundation
through the grants 200021-101924/1 and 200020-105260/1.

\vfill
{\em Note added in proof ---}  
After our paper appeared on the astro-ph, a second, similar study was posted there 
(McConnachie \& Irwin 2005, MNRAS, accepted, astro-ph/0510654). 
These authors analyzed the distribution of M31's satellite distribution in a coordinate system 
similar to the system presented here.
They also claim the existence of numerous possible candidate streams, and 
suggest that most of these streams may be chance alignments. 

\clearpage

\begin{figure}
\plottwo{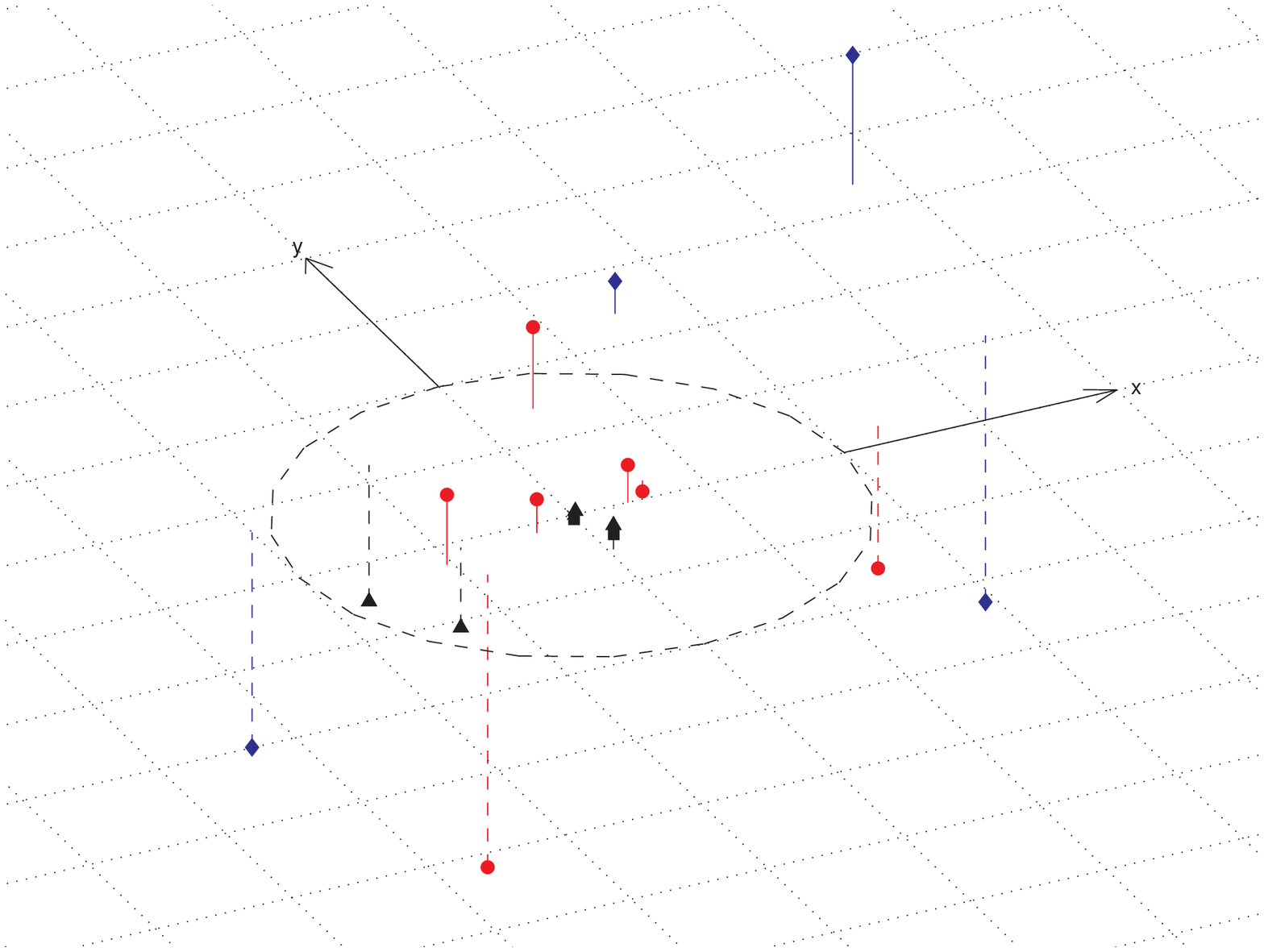}{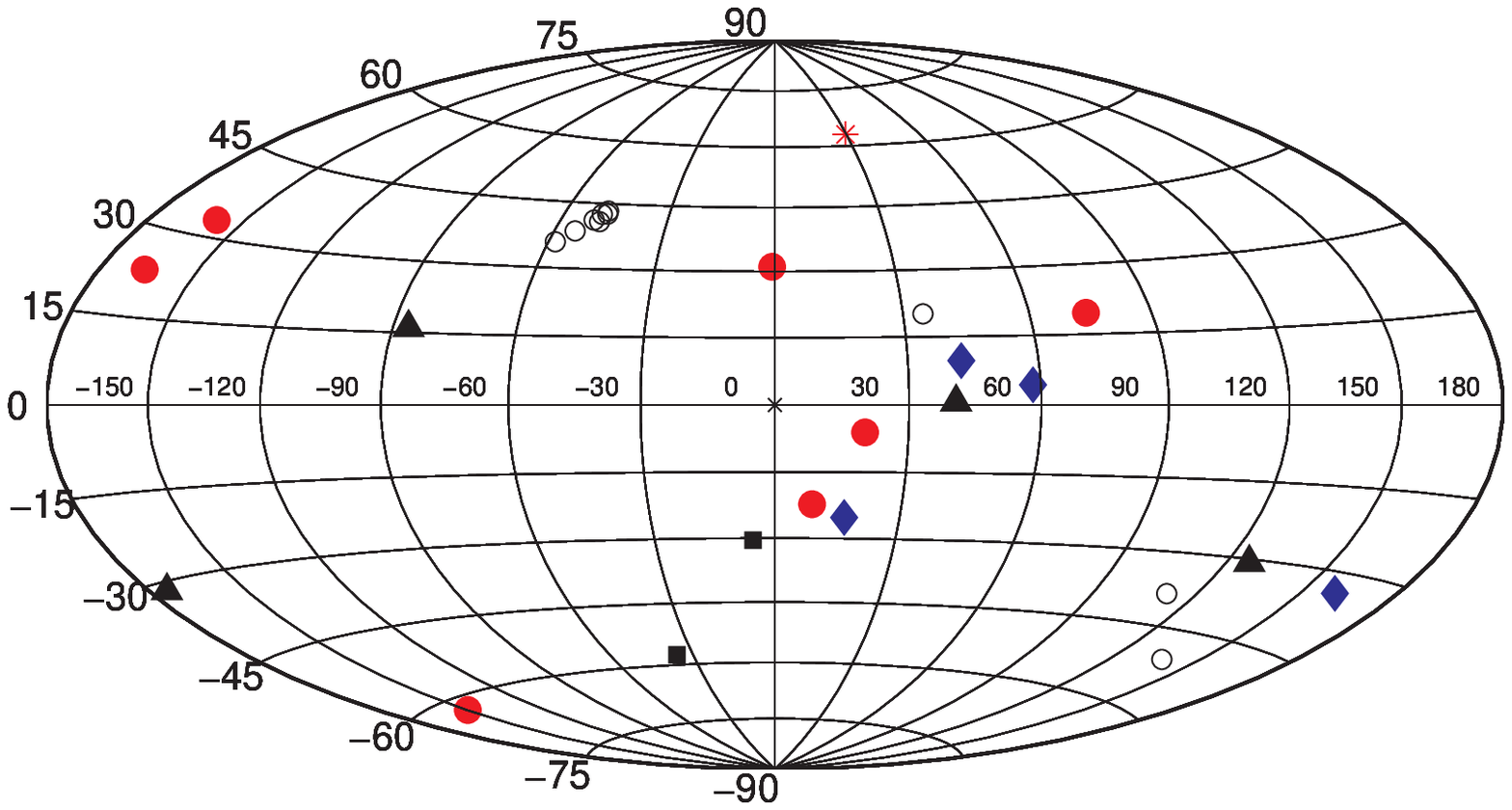}
\caption{Illustration of the position of the M31 satellites relative
to the disk plane of M31 (left panel).  The dotted grid indicates the
location of M31's disk plane, which contains the x- and y-coordinates
of our coordinate system centered on M31.  Solid (dashed) lines
indicate companion galaxies above (below) this plane.  The different
symbols refer to the morphological types of the M31 companions: dSphs
(filled red circles), dEs and cEs (filled black triangles), and dIrrs and
dIrr/dSphs (filled blue diamonds). M31 itself is marked with a cross.  The
axes of each grid have a length of 100\,kpc.  The dashed circle
circumscribes the central 200\,kpc around M31.  --- The right panel shows an
Aitoff projection of the same data in the M31 reference system, also
including the Andromeda stream (open circles), 
 its two most massive 
globular clusters (black filled squares) 
and M\,33 (asterisk).
Note the lack of an obvious Holmberg effect.  Visual inspection
suggests a possible great circle of satellites along the latitudes of
approximately $+30\degr$ and $-150\degr$. (See the electronic edition of 
the Journal for a color version of this figure.)} 
\end{figure}

\begin{figure}
\plotone{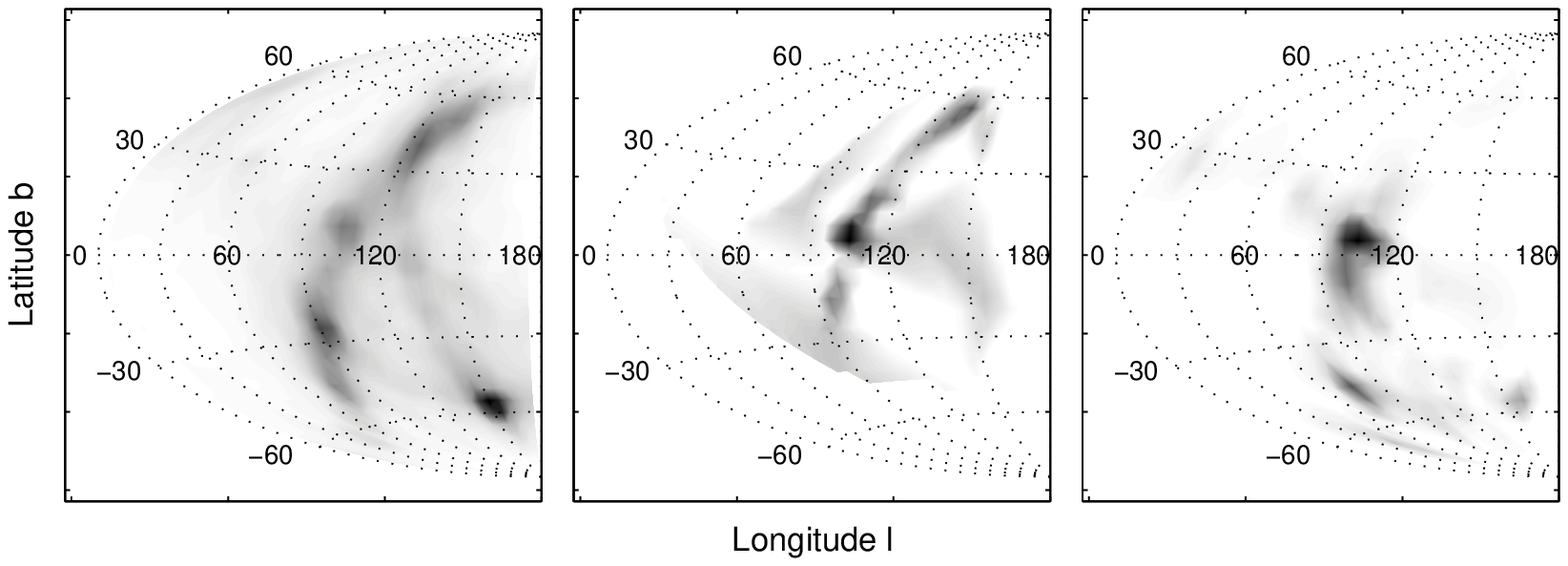}
\plotone{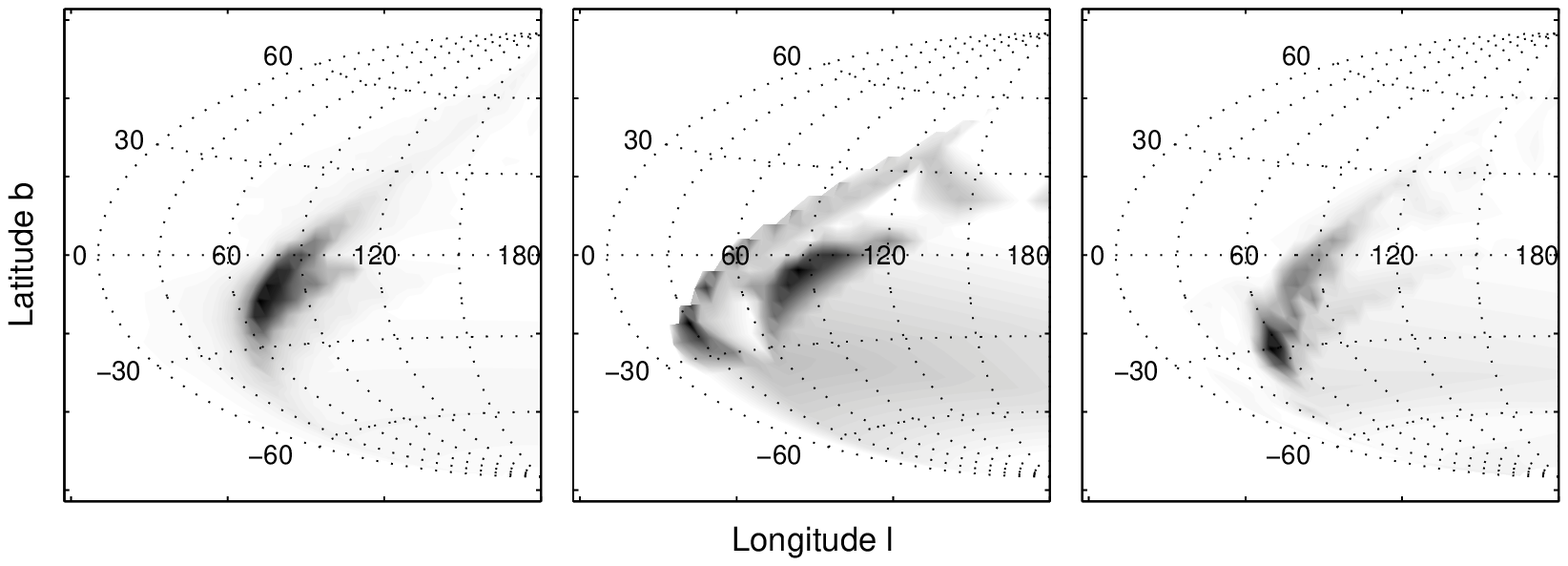}
\caption{Number density distributions of the normal vectors from all
bootstrap runs. The top line is drawn from the fits to the observed galaxy sample, 
whereas the bottom plots show {\em one} sample each of the large number of tests run 
on a random distribution. 
The left panels refer to all possible fits of great
planes to galaxies from the entire sample of the 15 M\,31 companions.
The middle panel shows results for fitting only the seven dSphs.  The
distribution after exclusion of the And\,II dSph and inclusion of the
dEs and M32 is shown in the right panel.  Distinct maxima  in the observed plots at
($l\,=\,150.8\degr$, $b\,=\,-56.4\degr$), ($l\,=\,107.1\degr$,
$b\,=\,6.9\degr$) and ($l\,=\,102.6\degr$, $b\,=\,5.2\degr$) indicate
poles of the respective best-fit planes.  Maxima in the random distribution do not stand out 
clearly and appear smeared out.} 
\end{figure}

\begin{figure}
\plotone{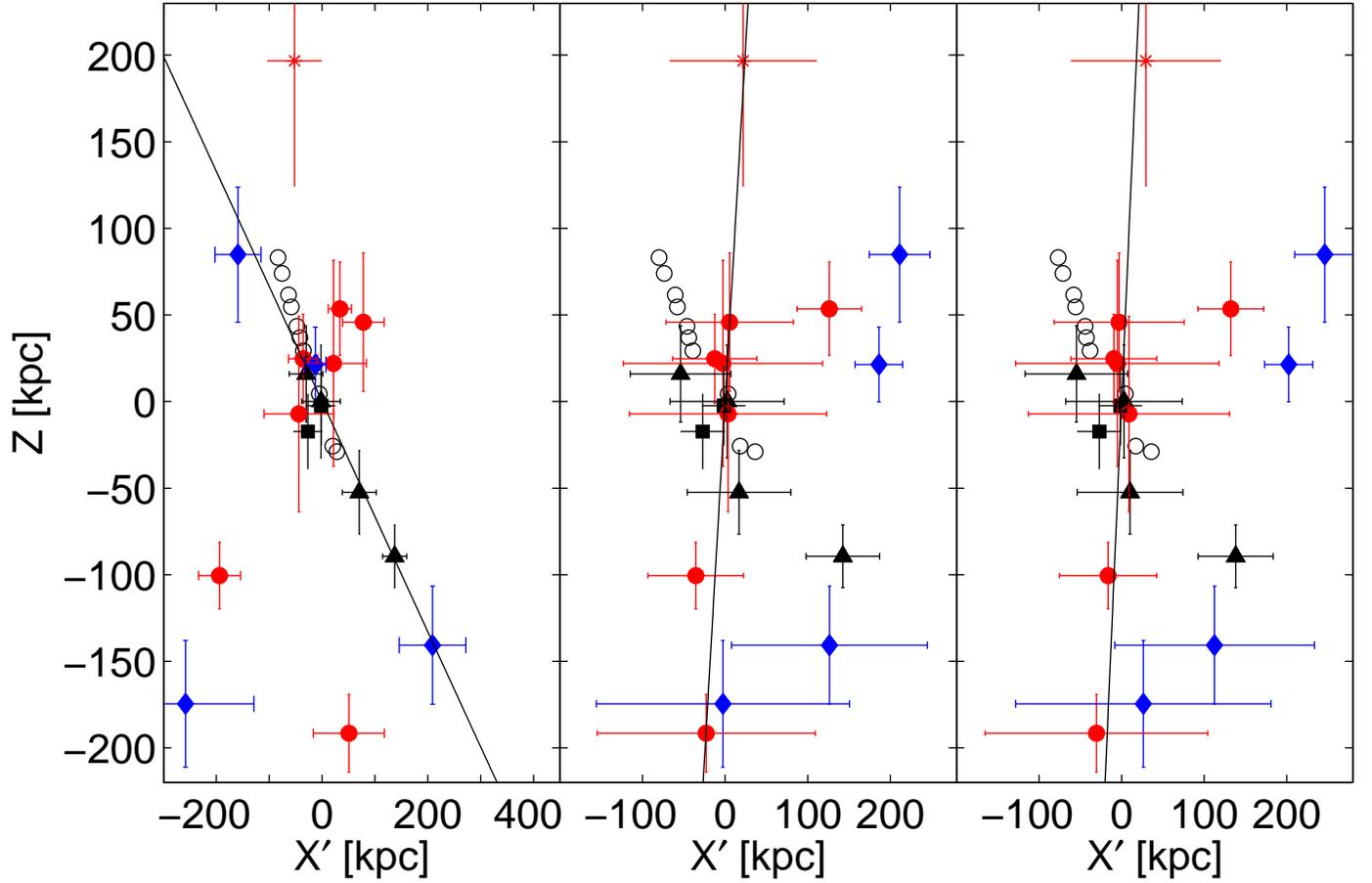}
\caption{The position of the satellite galaxies shown in edge-on
projections perpendicular to the best-fit planes.  The left plot shows
the fit to the entire dwarf sample. The middle panel illustrates the
best fit to the dSph subsample.  The right panel displays the rotated
CS and incorporates the best fit to the combined dE/cE and dSph sample
while excluding the outlier And\,II. The symbols are the same as in
Fig.~1. Note that the horizontal errorbars in these projections
indicate the combined uncertainties of the $X_{M31}$ and $Y_{M31}$
positions. (See the electronic edition of 
the Journal for a color version of this figure.)} 
\end{figure}

\begin{figure}
\begin{center}
\includegraphics[height=10cm]{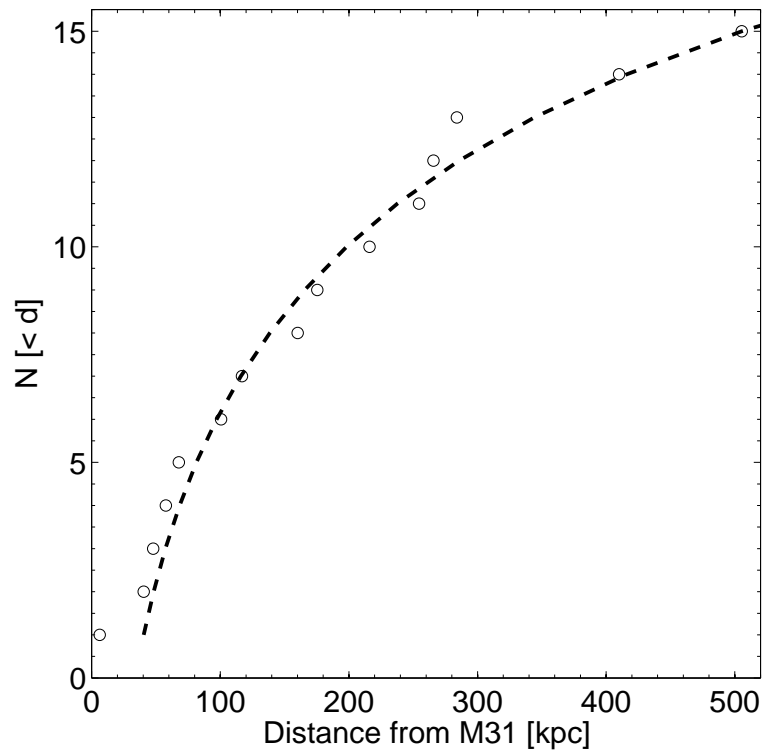}
\end{center}
\caption{Cumulative radial distribution of the M31 satellites. The dashed line is 
a power-law with an exponent of $-2$.} 
\end{figure}

\begin{figure}
\includegraphics[height=7cm]{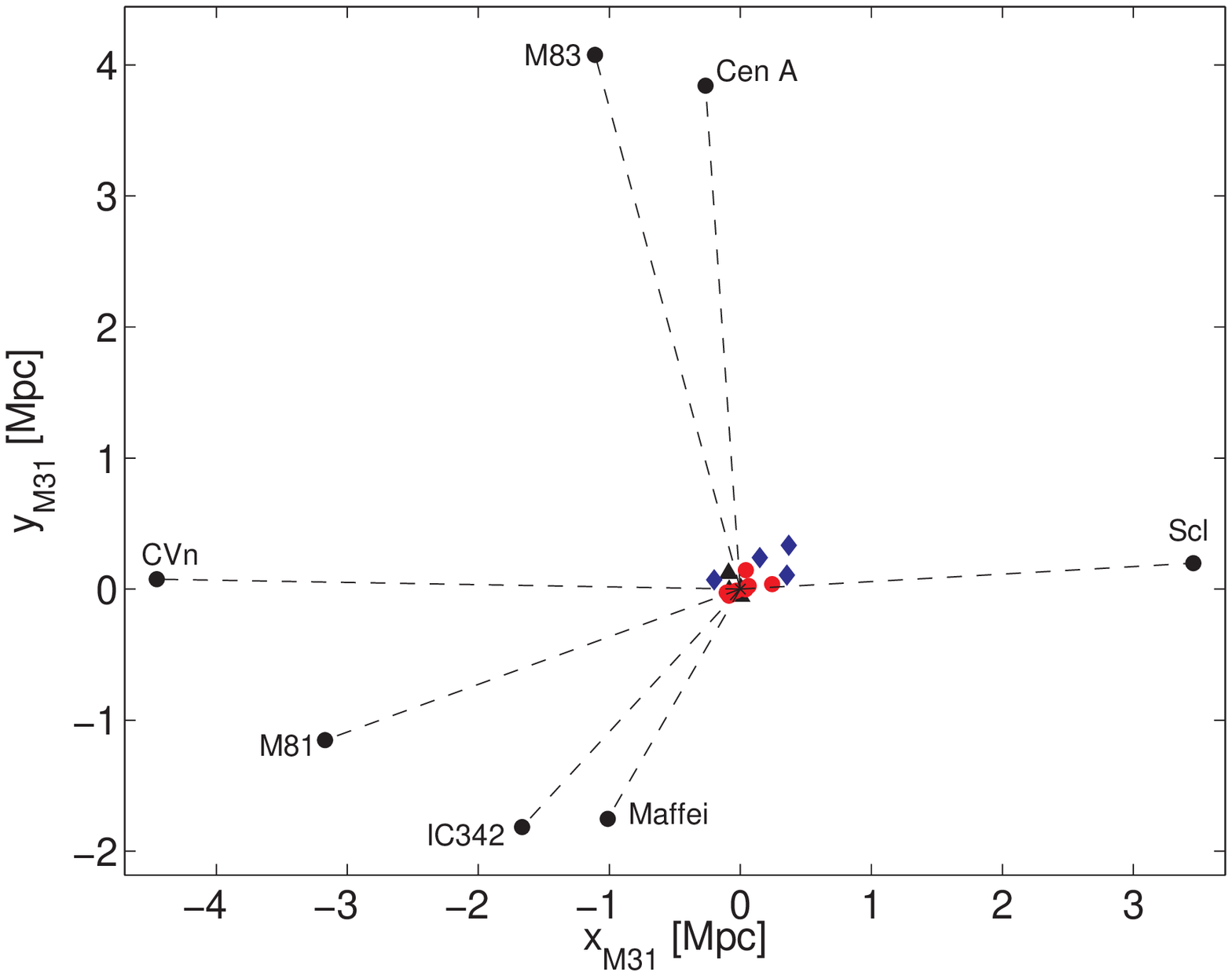}
\hfill
\includegraphics[height=7cm]{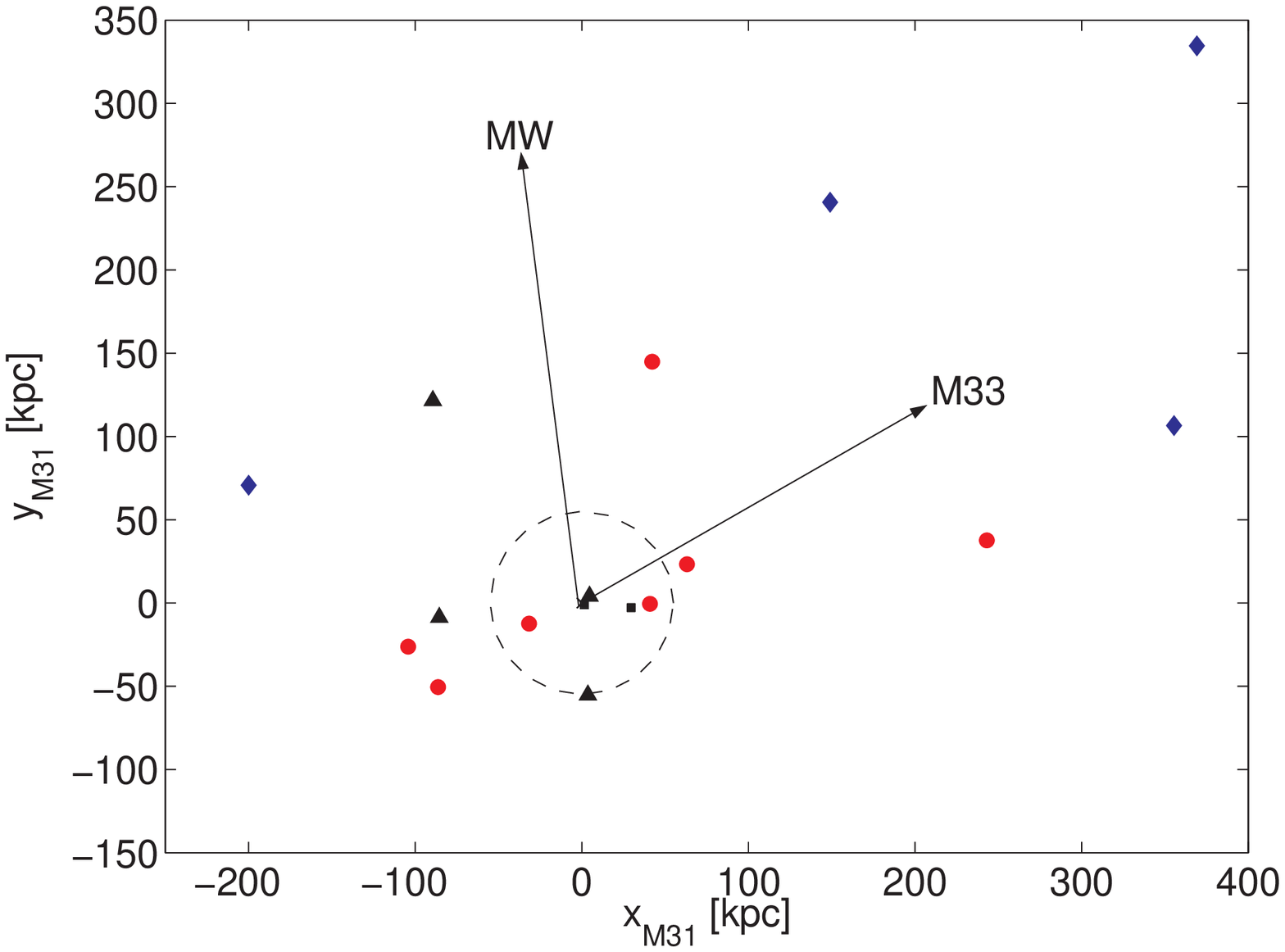}
\caption{Face-on views onto M31's disk plane. The left panel shows the
projected location of nearby galaxy groups as given by their most
luminous member. The right panel is a zoom on M31's immediate
vicinity, showing its satellites.  The circle designates the central
55 kpc, corresponding to the optical radius of M31's disk.  Arrows
indicate the direction of the Milky Way (MW) and M33. The symbols are
as in Fig.~1.  The polar great plane of M31's early-type satellites
lies along an axis pointing toward the M81 group (left panel) and
toward M33 (right panel). (See the electronic edition of 
the Journal for a color version of this figure.)} 
\end{figure}

\clearpage

\begin{table}
\begin{center}
\caption{Satellite sample. Coordinates and heliocentric distances from 
Rich et al.\ (1996), Grebel (2000), Barmby et al.\ (2002), Grebel et al.\ (2003), Zucker et al.\ (2004), Harbeck et 
al.\ (2005).  X, Y, and Z coordinates are given in our Cartesian
coordinate system centered on M31. }
\begin{tabular}{llcccrrr}
\hline
\hline
Galaxy & Type & $\alpha$ (J2000) &
$\delta$ (J2000) &
d$_{\odot}$ [kpc] & X$_{M31}$ [kpc] & Y$_{M31}$ [kpc] & Z$_{M31}$ [kpc]\\
\hline
M\,31   &       Spiral          & 00 42 44      & +41 16 09     & 773$\,\pm\,$20 &          0.0 &        0.0 &     0.0 \\
M\,32   &       cE              & 00 42 42      & +40 51 55     & 770$\,\pm\,$40 &          4.7 &        4.0 &     0.1 \\
NGC\,205 &      dE              & 00 40 22      & +41 41 07     & 830$\,\pm\,$35 &          3.8 &     $-55.3$&    16.0 \\
And\,I &        dSph            & 00 45 40      & +38 02 28     & 790$\,\pm\,$30 &         41.0 &      $-0.5$&    24.7 \\
And\,III &      dSph            & 00 35 34      & +36 29 52     & 760$\,\pm\,$70 &         63.2 &       23.2 &   $-7.2$\\
NGC\,147 &      dE              & 00 33 12      & +48 30 29     & 755$\,\pm\,$35 &       $-85.5$&      $-8.7$&  $-52.4$\\
And\,V &        dSph            & 01 10 17      & +47 37 41     & 810$\,\pm\,$45 &      $-104.2$&     $-26.3$&    45.8 \\
And\,II &       dSph            & 01 16 30      & +33 25 09     & 680$\,\pm\,$25 &         42.2 &      144.9 &    53.5 \\
NGC\,185 &      dE              & 00 38 58      & +48 20 12     & 620$\,\pm\,$25 &       $-89.3$&      121.6 &  $-89.4$\\
Cas\,dSph &     dSph            & 23 26 31      & +50 41 31     & 760$\,\pm\,$70 &       $-86.3$&     $-50.5$& $-191.5$\\
IC\,10 &        dIrr            & 00 20 17      & +59 18 14     & 660$\,\pm\,$65 &      $-200.0$&       70.7 & $-140.7$\\
And\,VI &       dSph            & 23 51 46      & +24 34 57     & 775$\,\pm\,$35 &        243.1 &       37.6 & $-100.5$\\
LGS\,3 &        dIrr/dSph       & 01 03 53      & +21 53 05     & 620$\,\pm\,$20 &        149.1 &      240.6 &    21.4 \\
Peg\,DIG &      dIrr/dSph       & 23 28 36      & +14 44 35     & 760$\,\pm\,$100 &       355.5 &      106.5 & $-174.5$\\
IC\,1613 &      dIrr            & 01 04 47      & +02 07 02     & 715$\,\pm\,$35 &        369.2 &      334.5 &    84.8 \\
And\,IX &       dSph            & 00 52 53      & +43 12 00     & 790$\,\pm\,$70 &       $-31.6$&     $-12.4$&    22.0 \\
M\,33   &       Spiral          & 01 33 51      & +30 39 37     & 847$\,\pm\,$60 &         87.4 &       49.8 &   196.7 \\
G\,1    & Globular cluster & 00 32 47     & +39 34 40     & 773$\,\pm\,$20 & 29.4 & $-2.8$ & $-17.4$ \\
B\,327  &	Globular cluster & 00 41 35     & +41 14 55     & 773$\,\pm\,$20 & 1.3 & $-0.9$ & $-2.5$ \\	
\hline
\end{tabular}
\end{center}
\end{table}

\begin{table}
\begin{center}
\caption{Effects of varying M31's distance on the resultant best-fit planes.}
\begin{tabular}{lcclc}
\hline
\hline
Fit sample & Adopted distance to M31 & \multicolumn{2}{c}{Best-fit plane} & Significance \\
 \mbox{ }          & [kpc]                   & \multicolumn{2}{c}{$(l,b)$} & \mbox{ } \\ 
\hline
                               & 760$\,\pm\,$20 & $150\fdg 7$&$-56\fdg 5$            & 86.9 \% \\
All satellites (15)            & 773$\,\pm\,$20 & $150\fdg 8$&$-56\fdg 4$ & 88.0 \% \\
                               & 783$\,\pm\,$20 & $150\fdg 2$&$-56\fdg 2$            & 87.1 \% \\
\hline
                               & 760$\,\pm\,$20 & $100\fdg 3$&\hspace{1.5ex}$10\fdg 9$       & 99.1 \% \\
All dSphs (7)                  & 773$\,\pm\,$20 & $107\fdg 1$&\hspace{3ex}$6\fdg 9 $ & 99.7 \% \\
                               & 783$\,\pm\,$20 & $101\fdg 9$&\hspace{3ex}$6\fdg 5 $           & 98.6 \% \\
\hline
                               & 760$\,\pm\,$20 & $102\fdg 7$&\hspace{2ex}$12\fdg 1$                 &  97.8 \% \\
dSphs (without And II),        & 773$\,\pm\,$20 & $102\fdg 6$&\hspace{3ex}$5\fdg 2 $ & 98.7 \% \\
 dEs, and M32 (10)             & 783$\,\pm\,$20 & $102\fdg 8$&\hspace{3ex}$4\fdg 8 $  & 99.2 \% \\
\hline
\end{tabular}
\end{center}
\end{table}

\end{document}